\documentclass{PoS}

\newcommand{\kms}{km s$^{-1}$}
\newcommand{\cmN}{cm$^{-2}$}
\newcommand{\cmn}{cm$^{-3}$}

\newcommand{\lam}{$\lambda$}

\newcommand{\civ}{\mbox{C\,{\sc iv}}}

\newcommand{\nv}{\mbox{N\,{\sc v}}}

\newcommand{\ovi}{\mbox{O\,{\sc vi}}}

\newcommand{\neviii}{\mbox{Ne\,{\sc viii}}}
\newcommand{\pv}{\mbox{P\,{\sc v}}}

\title{The Physics and Physical Conditions of Quasar Outflows}

\ShortTitle{Quasar Outflows}

\author{\speaker{Fred Hamann}$^a$, Daniel Capellupo$^b$, George Chartas$^c$, Sean McGraw$^{d}$, Paola Rodriguez Hidalgo$^{e}$, Joseph Shields$^{d}$, Jane Charlton$^{f}$, Michael Eracleous$^{f}$\\\
\llap{$^a$}University of Florida, Gainesville, Florida 32611, USA; \ E-mail:\email{fhamann@ufl.edu}\\
\llap{$^b$}Tel Aviv University, Tel Aviv, Israel\\
\llap{$^c$}College of Charleston, Charleston, South Carolina, USA\\
\llap{$^d$}Ohio University, Athens, Ohio, USA\\
\llap{$^e$}York University, Toronto, Canada\\
\llap{$^f$}Penn State University, State College, Pennsylvania, USA\\
}

\abstract{We describe two studies designed to characterize the total column densities, kinetic energies, and acceleration physics of broad absorption line (BAL) outflows in quasars. The first study uses new {\it Chandra} X-ray and ground-based rest-frame UV observations of 7 quasars with mini-BALs at extreme high speeds, in the range 0.1$c$ to 0.2$c$, to test the idea that strong radiative shielding (and therefore strong X-ray absorption) is needed to moderate the mini-BAL ionizations and facilitate their acceleration to extreme speeds. We find that the X-ray absorption is weak or absent, with generally $N_H < {\rm few} \times 10^{22}$ \cmN , and that radiative shielding is not important. We argue that the mini-BAL ionizations are controlled, instead, by high gas densities of order $n_H\sim 4\times 10^8$ \cmn\ in small outflow substructures. If we conservatively assume that the total column density in the mini-BAL gas is $N_H < 10^{22}$ \cmN , covering $>$15\% of the UV continuum source along our lines of sight (based on measured line depths), then the radial thickness of these outflows is only $\Delta R < 3\times 10^{13}$ cm and their transverse size is $>$$8\times 10^{15}$ cm. Thus the outflow regions have the shape of very thin ``pancakes" viewed face-on, or they occupy larger volumes like a spray of dense cloudlets with a very small volume filling factor. We speculate that this situation (with ineffective shielding and small dense outflow substructures) applies to most quasar outflows, including BALs. Our second study focuses from BALs of low-abundance ions, mainly PV 1118,1128 \AA , whose significant strengths imply large column densities, $N_H > 10^{22}$ \cmN , that can further challenge models of the outflow acceleration. In spite of the difficulties of finding this line in the Ly$\alpha$ forest, a search through the SDSS DR9 quasar catalog reveals $>$50 BAL sources at redshifts $z>2.3$ with strong PV BALs, which we are now using to characterize the general properties of high-column outflows. 
}

\FullConference{The Extreme sky: Sampling the Universe above 10 keV - extremesky2009,\\
		November 6-8, 2012\\
		Max-Planck-Insitut f\"ur Radioastronomie (MPIfR), Bonn,  			Germany}

\begin{document}

\section{Introduction}

Quasar outflows revealed by blueshifted broad absorption lines (BALs) are an important part of the quasar phenomenon. They are believed to be accelerated out from the atmospheres of quasar accretion disks by radiative forces, reaching observed speeds of a few thousand to tens of thousands of \kms . While substantial progress has been made in our understanding of BAL outflows, important questions still remain their basic properties and acceleration physics. For example, it has long been known that the intense radiation available to drive quasar outflows can also over-ionize them and make them too transparent for radiative driving. This problem appeared to be solved by models that invoke a highly-ionized and radiatively thick ``shield'' at the base of the flows to protect the BAL gas from over-ionization and facilitate its acceleration to high speeds \cite{Murray95}. These models are  supported by observations showing that BAL quasars are heavily absorbed in X-rays \cite{Gallagher02,Gallagher06}. However, this shielding picture runs into difficulty when we consider that the narrower cousins of BALs, the so-called mini-BALs, are not accompanied by strong X-ray absorption. They reach the same speeds with the same degrees of ionization as BALs {\it without} the benefits of a radiative shield. 

In \S2, we describe new observations and analyses of 7 mini-BAL quasars with extreme outflow speeds in the range 0.1$c$ to 0.2$c$ \cite{Hamann13}. These are highly successful flows that must have favorable conditions for their acceleration. Our main goal is to test the hypothesis that high outflow speeds require a strong radiative shield. We show that the radiative shielding is negligible in these sources and, therefore, it is not important for the acceleration. 

Another lingering problem is that measurements of the outflow column densities and kinetic energies are hampered by absorbing regions that only partially covering the background continuum source along our lines of sight. This can lead to absorption lines that do not reach zero intensity even if the line optical depths are orders of magnitude above unity. Section 3 presents early results from a program to find and measure BALs of rare ions, mainly PV 1118,1128 \AA , that are signatures of large total column densities, $N_H> 10^{22}$ \cmN , and extreme saturation in more commonly measured BALs like \civ\ \lam\lam1548,1551  and \ovi\ \lam\lam 1032,1038 \cite{Capellupo13}. 

\section{Radiative Shielding in High-Velocity Mini-BAL Outflows}

We selected 7 mini-BAL quasars with extreme outflow speeds ($v> 38,000$ \kms ) from our catalog of quasar outflow lines in the SDSS \cite{Paola08}. Figure 1 compares new rest-frame UV spectra obtained at the MDM observatory to earlier data from the SDSS or Lick Observatory (2001-2003). {\it Chandra} X-ray observations were obtained roughly concurrent with the MDM data (circa 2011). We find that most of the mini-BALs varied (Figure 1). This is consistent with the standard picture of mini-BALs belonging to the same general outflow phenomenon as BALs, close to the central black hole/accretion disk. The {\it Chandra} observations reveal weak or negligible amounts of X-ray absorption, with $\Delta\alpha_{ox} = \alpha_{ox}({\rm observed}) - \alpha_{ox}({\rm expected})$ typically 0.0 to $-$0.1 and spectral fitting to the brightest individual source (also \cite{Paola11}) and joint fitting to the ensemble dataset  indicating total X-ray column densities $N_H<2\times 10^{22}$ \cmN\ of neutral-equivalent absorption. These results are consistent with previous studies of lower-velocity mini-BALs that find X-ray absorbing columns at least an order of magnitude less than typical BAL quasars \cite{Gibson09,Wu10}.

\begin{figure}
 \includegraphics[scale=0.86,angle=-90.0]{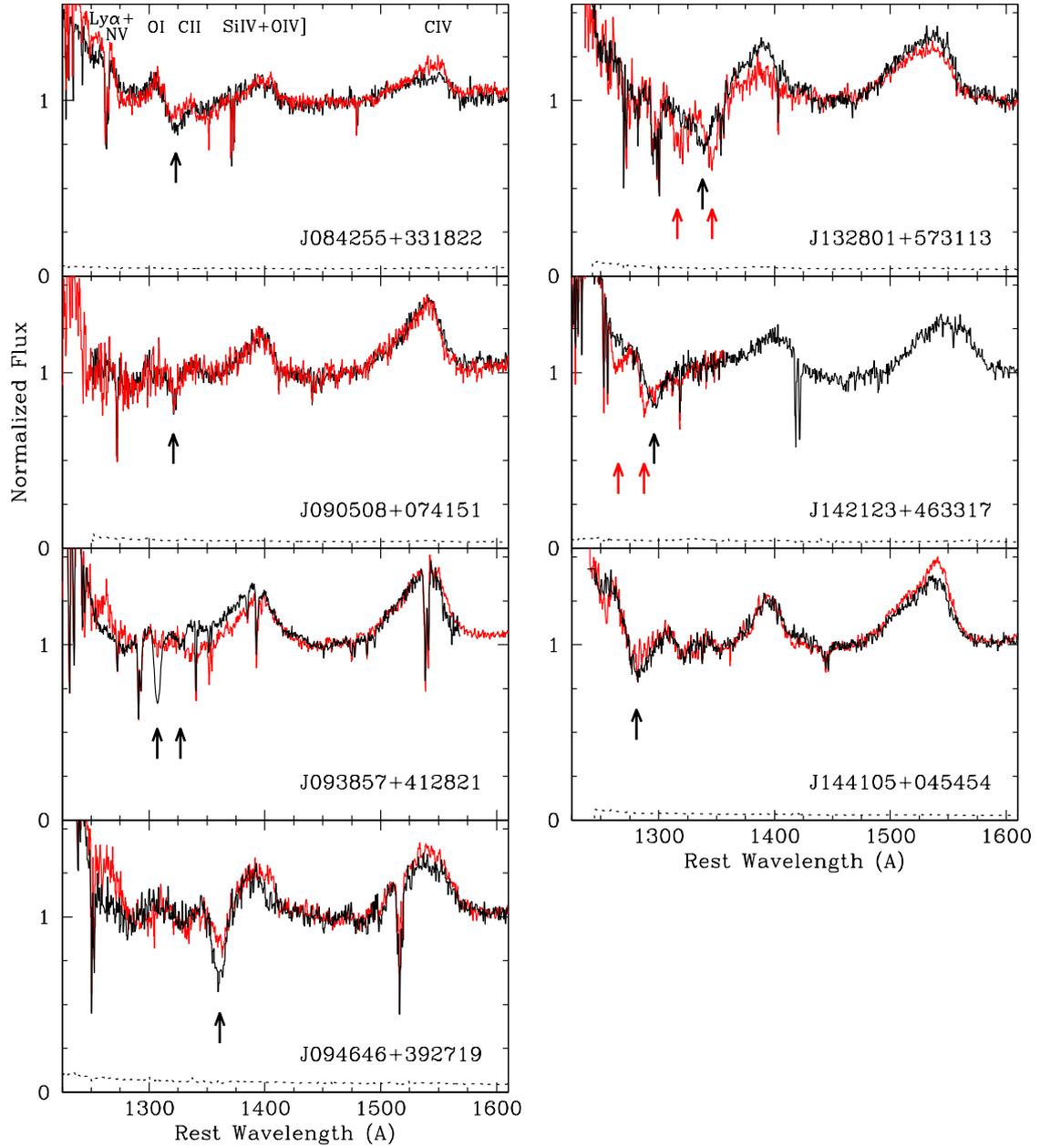}
\vspace{-10pt}
 \caption{Normalized MDM spectra in the quasar rest frames (red curves) overlaid on top of previous SDSS or Lick Observatory measurements (black curves) for the seven quasars in our mini-BAL sample. Significant mini-BALs are marked by red or black arrows in the MDM or previous spectra, respectively. Red arrows are not drawn if the mini-BAL disappeared or did not substantially change from the previous observation. The locations of several broad emission lines are shown across the top in the upper left panel.  }
\end{figure}

We performed extensive photoionization simulations with the spectral synthesis code Cloudy \cite{Ferland98} to determine the amounts of overall radiative shielding that might occur in these sources. These calculations are constrained by the observed weak X-ray absorption and weak or absent absorption lines in the UV (e.g., near velocity $\sim$ 0 where the shield is expected to reside). Table 1 and Figure 2 present the main results. The first three models listed in the table adopt a very large total column, $\log N_H = 23.5$ \cmN , and large doppler parameters, $b=100$ to 1000 \kms\ to maximize the shielding. The ionization parameter, $U$, is adjusted to produce optical depth 0.1 in the \civ\ or \ovi\ lines, consistent with the non-detections of these lines arising from the shield. These models can be ruled out because they produce too much X-ray absorption (Figure 1 and $\Delta\alpha_{ox}$ in Table 1). The fourth Cloudy model, called noC4b100xr, adopts a smaller total column density to be marginally consistent with both the UV and X-ray data. 

\begin{table}
 \centering
\begin{minipage}{151mm}
  \caption{Theoretical X-Ray Absorber/Shielding Results$^a$}
  \tabcolsep=1.82mm
  \begin{tabular}{@{}lccccccccccc@{}}
  \hline
  & & & & \multispan{4}{\hfil --- Line Optical Depths ---\hfil}& \multispan{2}{\hfil ------ HR ------\hfil}\\
 ~~Name & $\log N_H$ & $\log U$ & $b$ & \civ & \nv & \ovi & \neviii & z=2.0& z=3.3& ~$\Delta \alpha_{ox}$~\\ 
\hline
{\it Cloudy Models:}\\[-1.3pt]
~~noC4b100 & 23.5 & 2.11 & 100 & 0.1 & 2.5 & 226 & 365 & 4.8 & 2.9 & -0.64\\[-1.3pt]
~~noC4b1000 & 23.5 & 2.09 & 1000 & 0.1 & 0.9 & 36 & 56 & 5.1 & 3.0 & -0.71\\[-1.3pt]
~~noO6b1000 & 23.5 & 2.26 & 1000 & ... & ... & 0.1 & 3.5 & 3.0 & 2.4 & -0.43\\[-1.3pt]
~~noC4b1000xr & 22.9 & 1.60 & 1000 & 0.2 & 1.3 & 47 & 76 & 2.5 & 1.8 & -0.27\\
{\it Neutral Absorbers:}\\[-1.3pt]
~~neutral22 & 22.0 & ... & ... & ...& ...& ...& ...& 1.5 & 1.2 & -0.07\\[-1.3pt]
~~neutral22.5 & 22.5 & ... & ... & ...& ...& ...& ...& 2.3 & 1.6 & -0.23\\[-1.3pt]
~~neutral23 & 23.0 & ... & ... & ...& ...& ...& ...& 4.1 & 2.5 & -0.68\\
\hline
\end{tabular}
\small
$^a$The column densities, $N_H$, and doppler velocities, $b$, have units \cmN\ and \kms , respectively. The line optical depths apply to the short wavelength components of the doublets \civ\ \lam 1548, \nv\ \lam 1239, \ovi\ \lam 1032, and \neviii\ \lam 770. Hardness ratios, HR, are listed for redshifts $z=2.0$ and $z=3.3$. For comparison, the unabsorbed X-ray continuum used in these calculations, with $\alpha_x=-0.9$, has HR = 0.85. 
\end{minipage}
\end{table}

\begin{figure}
\includegraphics[scale=0.34,angle=-90.0]{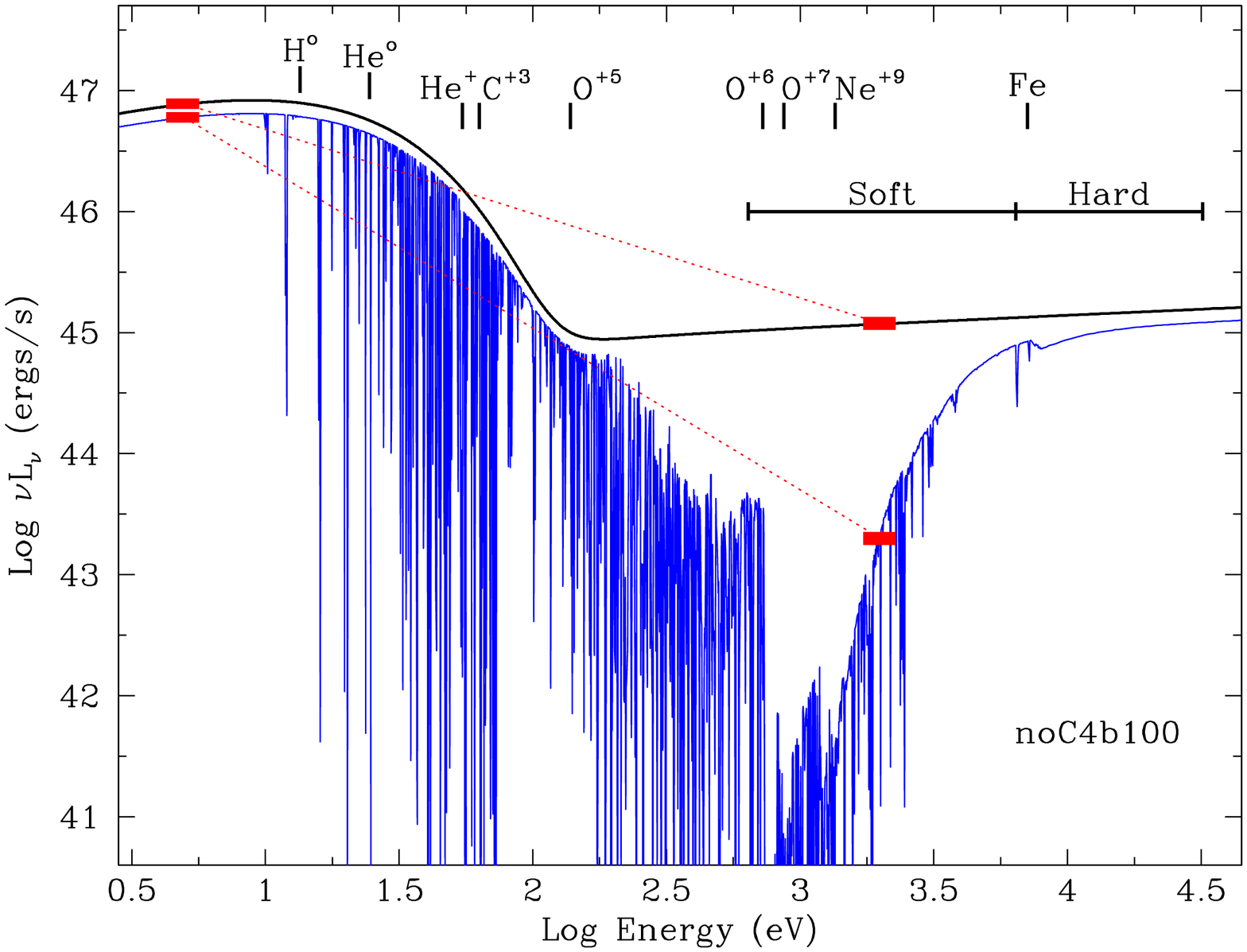}
\includegraphics[scale=0.34,angle=-90.0]{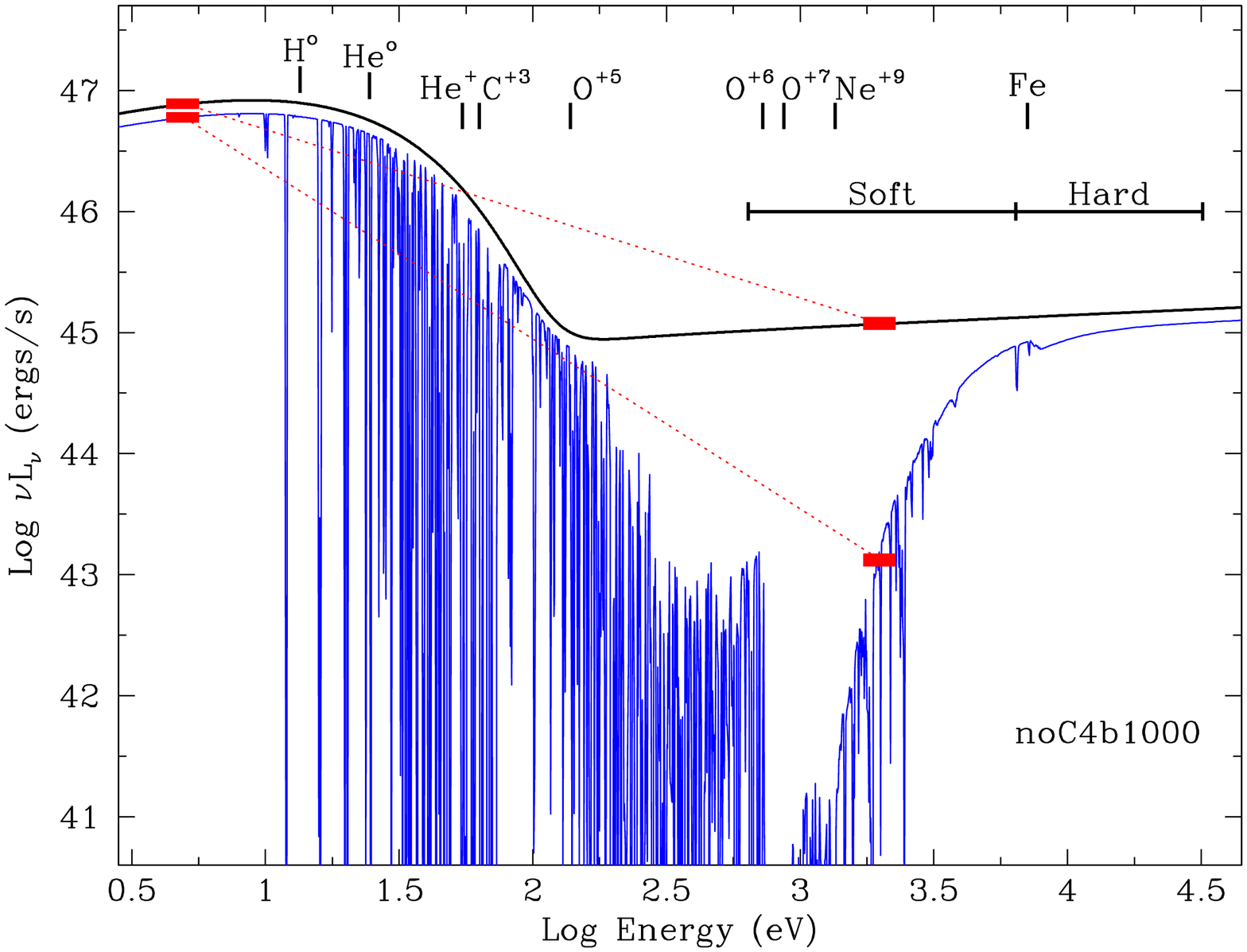}
\includegraphics[scale=0.34,angle=-90.0]{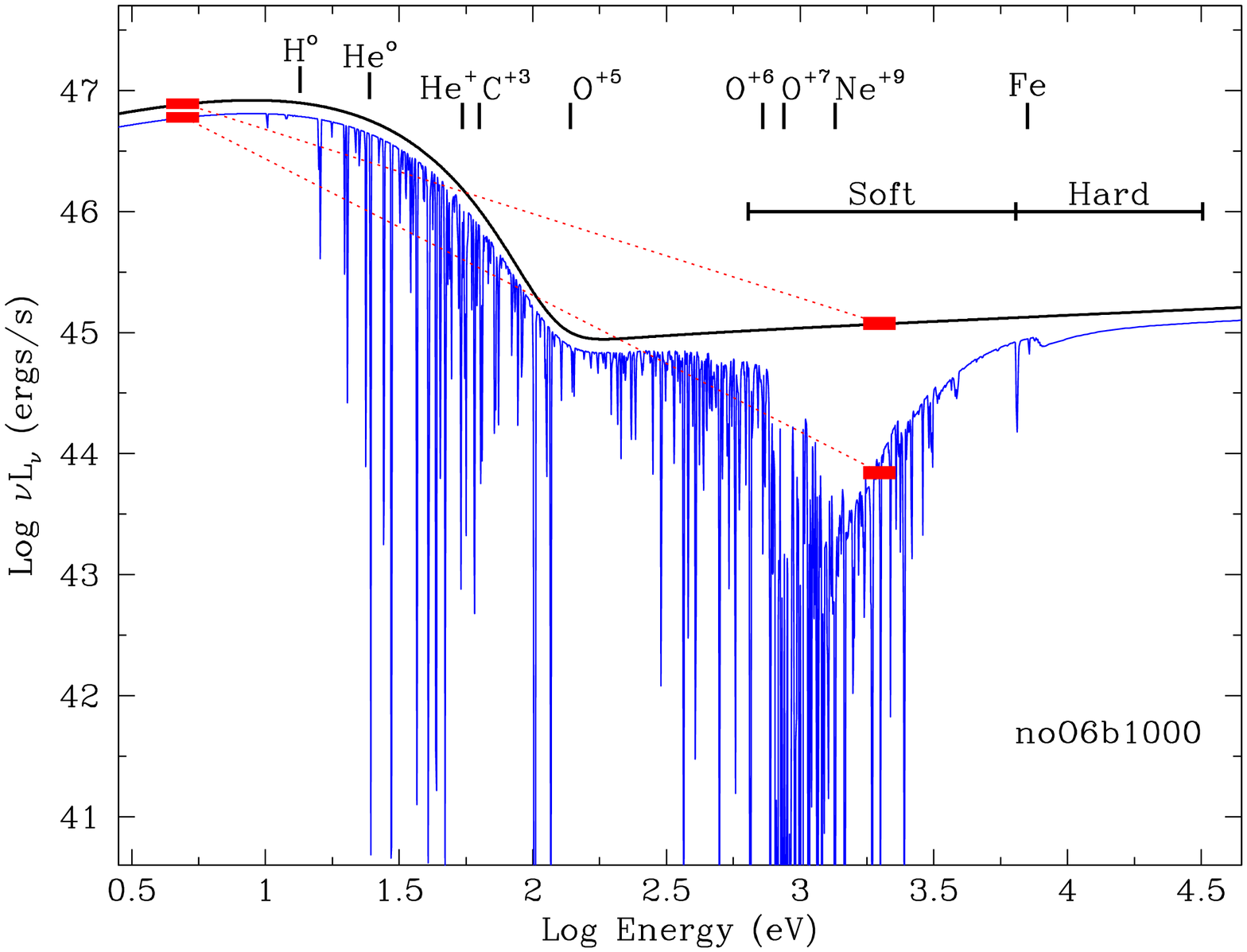}
\includegraphics[scale=0.34,angle=-90.0]{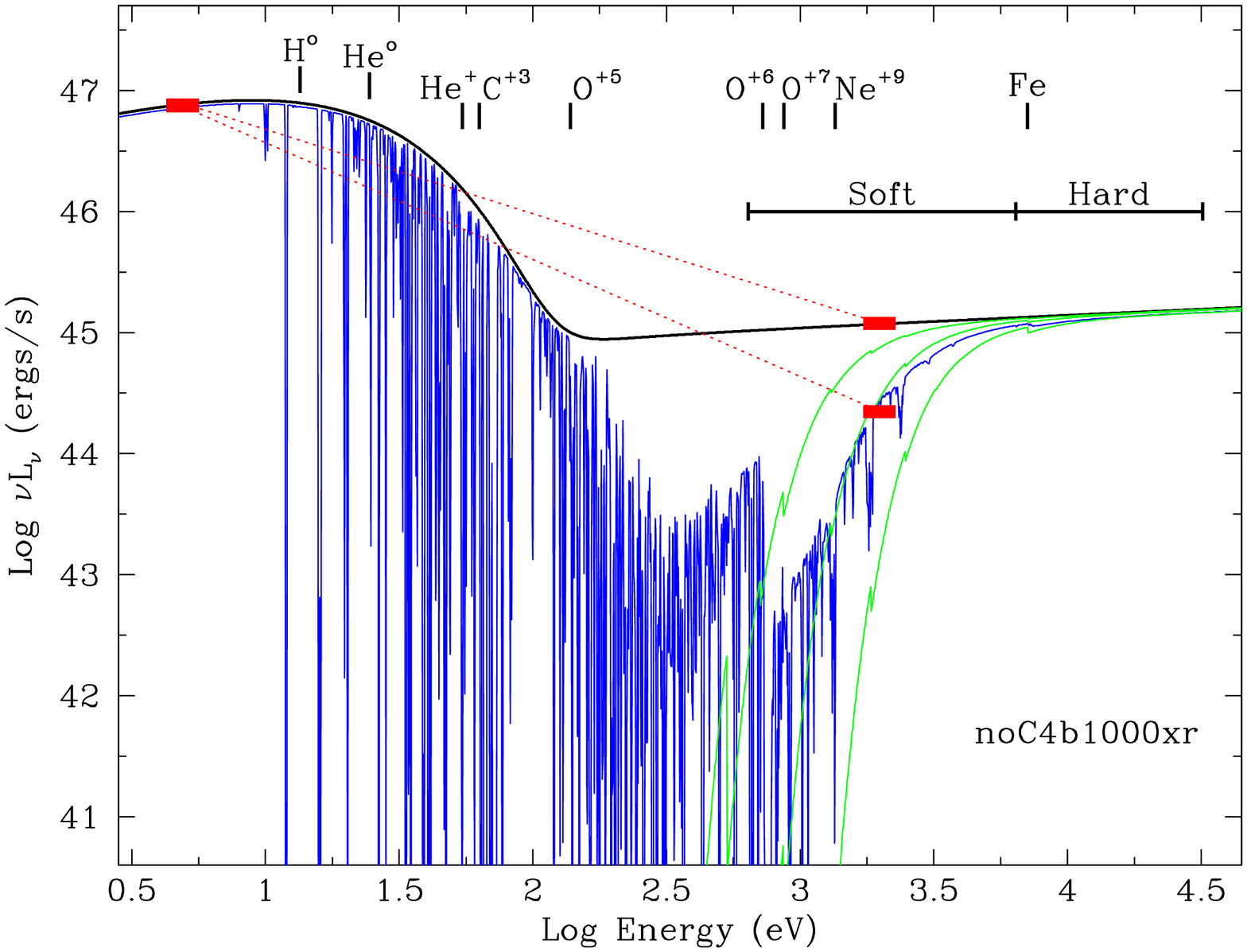}
\vspace{-3pt}
 \caption{Incident spectra (black curves) and transmitted spectra (blue curves) for the four Cloudy models listed in Table 1. The model names are given in the lower right of each panel. The bold red dashes connected by thin red dotted lines show the luminosities at 2500 \AA\ (4.96 eV) and 2 keV used to calculate $\alpha_{ox}$. Green curves in the lower right panel show transmitted X-ray spectra for the three neutral absorbers in Table 3, with $\log N_H = 22$, 22.5, and 23 \cmN\ from left to right. Ionization energies for some important ions are marked across the top. The observed Soft (0.2-2.0 keV) and Hard (2.0-10 keV) X-ray bands are shown by horizontal dashes for illustration at redshift $z=2.2$. Small offsets of the transmitted spectra below the incident spectrum are due to electron scattering.  }
\end{figure}

However, our main result is that these none of these quasars have significant radiative shielding because the putative absorbers do  not block the far-UV radiation that can destroy important ions like \civ\ and \ovi\ in the mini-BAL gas. This is evident from the weak absorption predicted near the ionization edges of these ions in Figure 1. Thus we find that while mini-BAL outflows reach the same speeds with the same moderate degrees of ionization as BAL outflows, they do so {\it without} the benefits of a radiative shield. We propose that the outflow ionizations are controlled, instead, by high gas densities defined by photoionization requirements to be
\begin{equation}
n_H \; = \; 4\times10^8 \left({{\nu l_\nu (2500{\rm \AA })}
\over{8\times 10^{46}\,{\rm ergs/s}}}\right)
\left({{0.4}\over{U}}\right)\left({{2\,{\rm pc}}\over{R}}\right)^2
~{\rm cm}^{-3}
\end{equation}
where the luminosity $\nu l_\nu (2500{\rm \AA }) = 8\times 10^{46}$ ergs s$^{-1}$ is roughly typical of our sample, $R = 2$ pc is a reasonable guess for the location of the mini-BAL gas, and $U=0.4$ is an ionization parameter consistent with strong \civ\ and \ovi\ mini-BALs. High gas densities then imply that the flows are like a fine spray, composed of many small substructures with a very small volume filling factor. These results support models like magnetic disk winds \cite{deKool95}, that can confine small clouds with magnetic pressure and do not require a radiative shield for the acceleration.

\section{PV \& Large Outflow Column Densities}

\pv\ \lam\lam 1118,1128 is an important diagnostic of large column densities in BAL outflows \cite{Hamann98,Borguet12}, but it can be difficult to measure in ground-based spectra because the short wavelengths require high redshift quasars that have severe Ly$\alpha$ forest contamination below 1216 \AA . We visually inspected $\sim$3000 BAL quasars in the SDSS DR9 with redshifts, $z>2.3$, high enough to measure \pv\ BALs. We find $>$50 quasars with definite strong \pv\ BALs and many more quasars with \pv\ BALs probably present. Figure 3 shows four quasars with strong \pv\ BALs. We are presently working to characterize the BAL and broad emission line properties of this high column density sample, e.g., compared to other BAL and non-BAL quasar samples, and derive basic constraints on the outflow properties from photoionization analyses of the relative BAL strengths \cite{Capellupo13}.

\begin{figure}
\begin{center}
 \includegraphics[scale=0.245,angle=0.0]{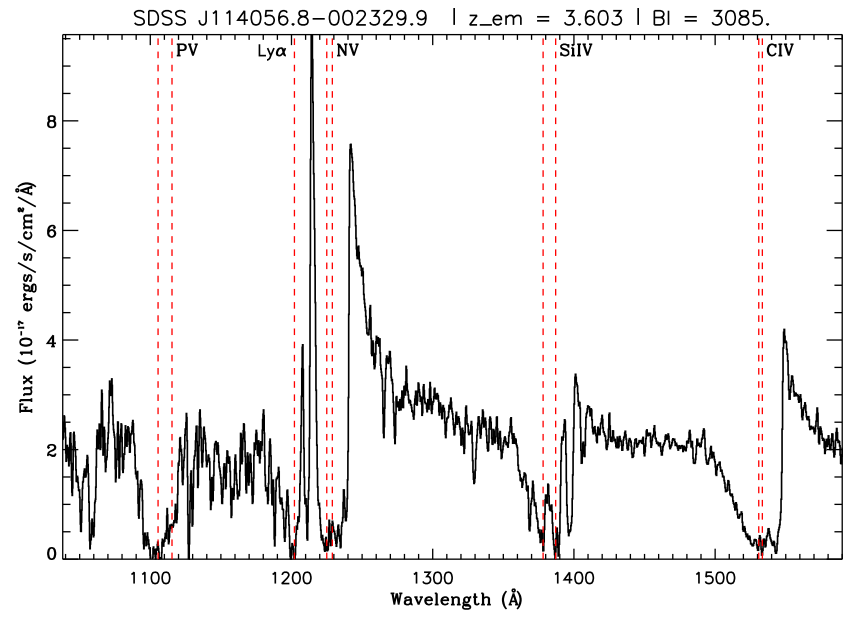}
 \includegraphics[scale=0.245,angle=0.0]{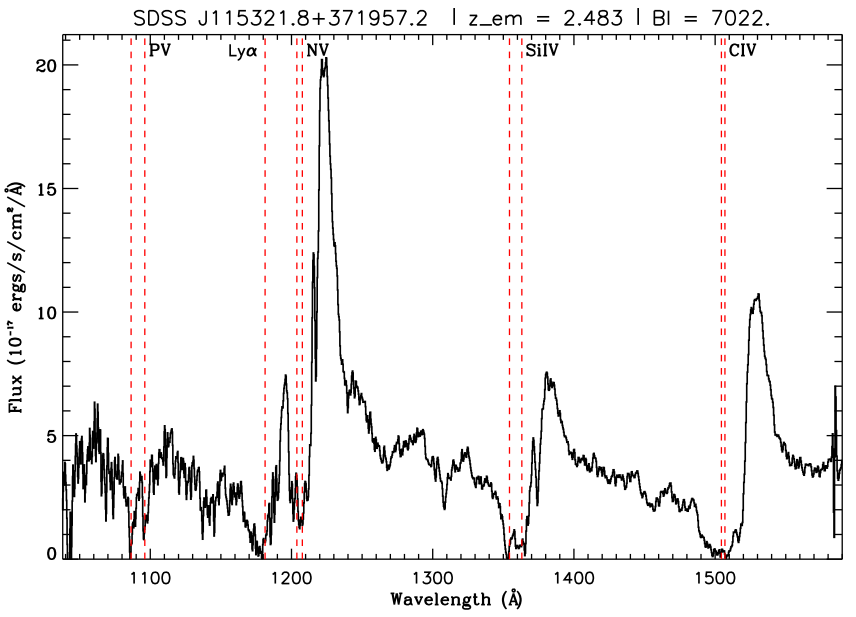}
 \includegraphics[scale=0.245,angle=0.0]{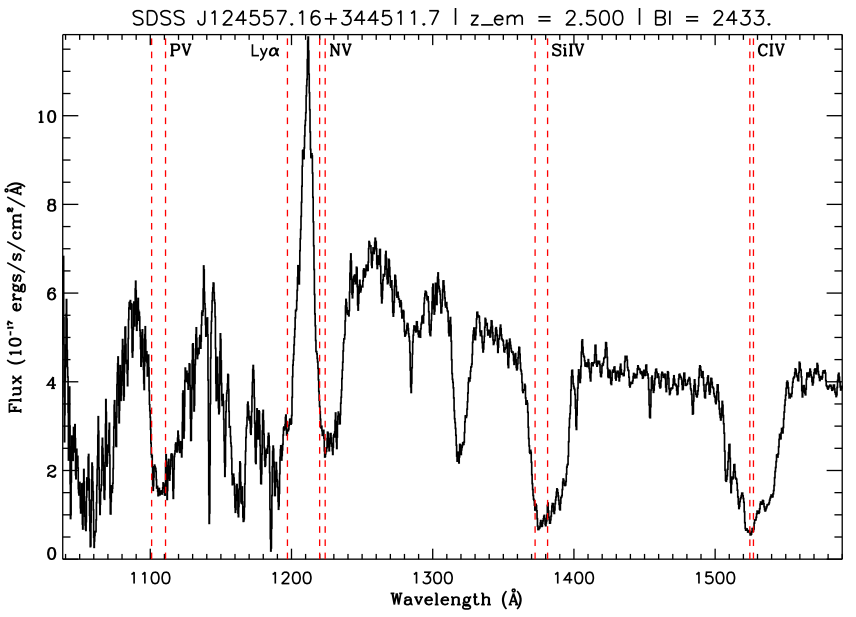}
 \includegraphics[scale=0.245,angle=0.0]{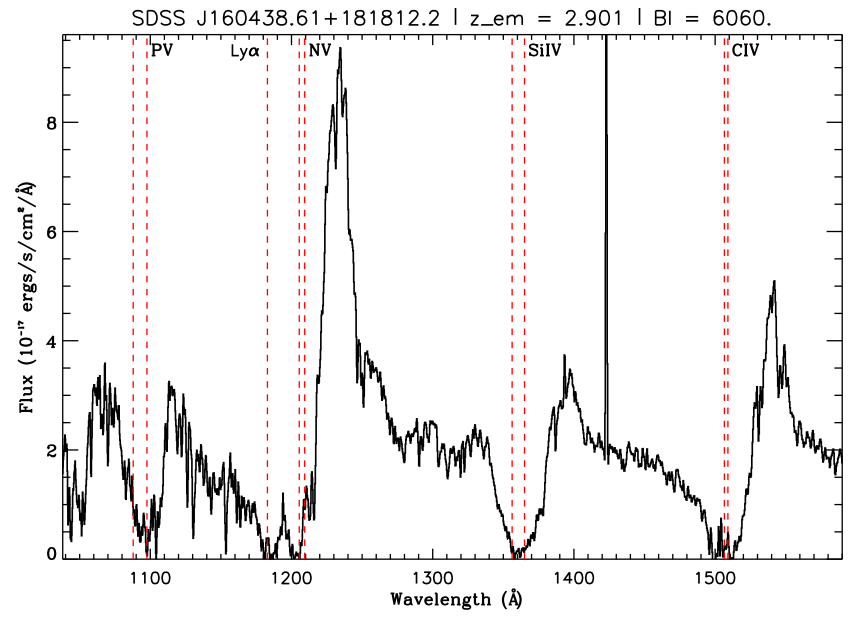}
\vspace{-10pt}
 \caption{Rest-frame UV spectra of quasars from the SDSS3 DR9 that are representative of our sample with strong PV absorption \cite{Capellupo13}. Several prominent BALs are labeled by red dashed vertical lines.}
   \end{center}
\end{figure}

\end{document}